\documentclass[a4paper,12pt]{article}
\usepackage[english]{babel}
\usepackage{amssymb} 
\usepackage{amsthm}
\usepackage[centertags]{amsmath}
\usepackage[latin1]{inputenc}
\usepackage{rotating}
\usepackage{graphicx,graphics}
\usepackage{enumerate}
\usepackage{multirow}
\usepackage{array}
\usepackage{physics}
\usepackage{siunitx}
\usepackage{subfigure}
\usepackage{pict2e}
\usepackage[T1]{fontenc}
\usepackage{scalefnt}
\usepackage{lscape}
\usepackage{blindtext}
\usepackage[linesnumbered,ruled,vlined]{algorithm2e}
\usepackage{bm}
\usepackage{booktabs,multirow,here}
\usepackage{enumitem,array}
\usepackage{thumbpdf,lmodern}
\usepackage[comma,authoryear]{natbib} 
\usepackage{color}
\usepackage[table,svgnames,dvipsnames]{xcolor}
\usepackage{listings}
\usepackage[noblocks]{authblk}
\usepackage{breakcites}
\usepackage[verbose]{placeins}
\usepackage[noblocks]{authblk}
\usepackage{setspace}
\usepackage{caption}
\usepackage{lineno}
\usepackage{ulem}
\usepackage{mathtools}
\usepackage[spaces,hyphens]{url}
\usepackage[pdfauthor={Author's name},%
pdftitle={Document Title},%
pagebackref=true,%
pdftex]{hyperref}

\usepackage[margin=2cm]{geometry}

\newcommand{\proglang}[1]{\textsf{#1}}

\usepackage{etoolbox}

\newcounter{algoline}

\AtBeginEnvironment{algorithm}{\setcounter{algoline}{0}}

\definecolor{codegreen}{rgb}{0,0.6,0}
\definecolor{codegray}{rgb}{0.5,0.5,0.5}
\definecolor{codepurple}{rgb}{0.58,0,0.82}
\definecolor{backcolour}{rgb}{0.95,0.95,0.92}

\definecolor{blue}{rgb}{0.0, 0.2, 0.6}
\definecolor{brown}{rgb}{0.65, 0.16, 0.16}

\lstdefinestyle{mystyle}{
    backgroundcolor=\color{backcolour},   
    commentstyle=\color{codegreen},
    keywordstyle=\color{magenta},
    numberstyle=\tiny\color{codegray},
    stringstyle=\color{codepurple},
    basicstyle=\footnotesize,
    breakatwhitespace=false,         
    breaklines=true,                 
    captionpos=b,                    
    keepspaces=true,                 
    numbers=left,                    
    numbersep=5pt,                  
    showspaces=false,                
    showstringspaces=false,
    showtabs=true,                  
    tabsize=2
}
 
\title{\bf Optimal sample size for the Birnbaum-Saunders distribution under a decision-theoretic approach}

\author{\small Eliardo G. Costa$^{1}$\thanks{Corresponding Author: Eliardo G. Costa.
Departamento de Estat\'istica, Universidade Federal do Rio Grande do Norte, Brazil. Email:
eliardocosta@ccet.ufrn.br}\,  and  Manoel Santos-Neto$^{2}$\\
{\footnotesize{$^1$Departamento de Estat\'{i}stica, Universidade Federal do Rio Grande do Norte, Brazil}}\\
{\footnotesize{$^2$Departamento de Estat\'istica, Universidade Federal de Campina Grande, Brazil}}}
\begin{document}
%\linenumbers
\maketitle

\begin{abstract}
The Birnbaum-Saunders distribution has been widely applied in several areas of science and although several methodologies related to this distribution have been proposed, the problem of determining the optimal sample size for estimating its mean has not yet been studied. For this purpose, we propose a methodology to determine the optimal sample size under a decision-theoretic approach. In this approach, we consider loss functions for point and interval inference. Finally, computational tools in the \proglang{R} language were developed to use in practice.\\

\end{abstract}

\section{Introduction}

\cite{bs69} introduced a family of distributions to model failure times for metals subject to periodic stress and provided a natural physical justification for this family. This family is the so-called Birnbaum-Saunders (BS) distribution. In the last decades, this distribution has received considerable attention in the literature and many methodologies have been proposed for parameter inference. Such attention is justified by its wide applicability and its variations have been applied in several areas including finance, business, engineering, environmental, medicine, quality control and many others. A detailed review of the BS distribution including methodologies under the frequentist and Bayesian approaches is presented in \cite{Balakrishnan2019}.

Although several methodologies related to this distribution have been proposed, the problem of determining an optimal size for estimating the mean of the BS distribution has not yet been studied. Recently, \cite{Bourguignonetal20} presented guidelines about the minimum sample size for monitoring the median parameter of the BS distribution in the context of quality control under a frequentist approach. In this way, we develop a methodology via a Bayesian decision-theoretic approach based on a criterion that minimizes the sum of the Bayes risk and the sampling cost. The proposed approach depends on an \textit{ad hoc} loss function defined to accommodate the implications of a decision. We consider four different loss functions for point and interval inference, two for each type of inference. Using the same approach but for other models there is a considerable literature, see for example \cite{EtzioniKadane93}, \cite{Sahu2006a}, \cite{Parmigiani2009}, \cite{Islam2012,IslamPettit14}, \cite{DeSantisGubbiotti16}, \cite{Costa17} and references therein.

The paper unfolds as follows. In Section~\ref{sec:2} we discuss the Bayesian model and the inference of the parameters of the BS distribution. In Section~\ref{sec:3} we present the methodology to obtain the optimal sample size under a decision-theoretic approach. Finally, we conclude with a discussion of the results in Section~\ref{sec:4}.

\section{Bayesian model}\label{sec:2}

Much of the information about the BS distribution presented in this section has been gathered from other works, for example, \cite{bs69,bs69b}, \cite{leivabook} and \cite{Balakrishnan2019}. Let $X$ be a BS distribution with a scale parameter $\beta$ and a shape parameter $\alpha$, we denote by $X \sim \mathrm{BS}(\alpha,\beta)$. Then, the respective probability density function is given by
\begin{equation*}
f_{X}(x|\alpha, \beta) =  
\frac{1}{\sqrt{2\,\pi}}\,\exp\left[-\frac{1}{2\alpha^2} \left(
\frac{x}{\beta}+ \frac{\beta}{x}-2\right)\right]
\frac{(x+\beta)}{2\alpha \sqrt{\beta\,x^{3}}}\qc x \in \real_{>0}\qc \alpha,\beta \in \real_{>0}.
\end{equation*}
Besides being the scale parameter, the parameter $\beta$ is also the median of this distribution. Furthermore, the mean and the variance of the BS distribution are given by 
\begin{equation}\label{eq:1}
  \theta\coloneqq\mathbb{E}[X] = \beta\left(1 + \frac{\alpha^2}{2} \right) \quad \text{and} \quad \mathrm{Var}[X] = (\alpha\beta)^2 \left(1 + \frac{5\alpha^2}{4} \right).
\end{equation}
Also, if $X$ is Birnbaum-Saunders distributed then
\begin{equation}\label{stoch-eq}
X = \frac{\beta}{4}\left(\alpha Z + \sqrt{(\alpha Z)^2 + 4}  \right)^2, 
\end{equation}
where $Z$ follows a standard normal distribution, which is useful to draw values from the $\mathrm{BS}(\alpha,\beta)$ distribution. Given a sample $\mathbf{x}_n=(x_1,\ldots,x_n)$, the likelihood function from the $\mathrm{BS}(\alpha,\beta)$ satisfies
\begin{equation*}%\label{likelihood}
 \mathcal{L}(\alpha,\beta;\mathbf{x}_n)\propto\frac{1}{(\alpha\beta)^n}\prod_{i=1}^n\left[\left(\frac{\beta}{x_i}\right)^{1/2}+\left(\frac{\beta}{x_i}\right)^{3/2}\right]\exp\left[-\frac{1}{2\alpha^2}\sum_{i=1}^n\left(\frac{x_i}{\beta}+\frac{\beta}{x_i}-2\right)\right].
\end{equation*}

For the parameters $\alpha$ and $\beta$ of the model, we consider proper prior distributions because the use of noninformative prior distributions yields an improper posterior distribution and continuous conjugate priors do not exist \citep{Wang2016}. A possible choice for a prior distribution for $\beta$ is the inverse gamma distribution in which the density satisfies
\begin{equation*}
\pi(\beta)\propto\beta^{-(a_1+1)}\exp\left(-\frac{b_1}{\beta}\right),\quad\beta \in \real_{>0}, 
\end{equation*}
where $a_1$ and $b_1$ are positive and known constants (hyperparameters), we denote by $\beta\sim\mathrm{IG}(a_1, b_1)$. We also assume a inverse gamma prior distribution for $\alpha^2$ with hyperparameters $a_2$ and $b_2$. Thus, we may write the model hierarchically as follows
\begin{eqnarray*}
&&X_i|\alpha,\beta\stackrel{\mathrm{iid}}{\sim} \mathrm{BS}(\alpha,\beta),\quad 
i=1,2,\ldots,n;\\
&&\beta\sim \mathrm{IG}(a_1,b_1)\quad\mathrm{and}\quad\alpha^2\sim \mathrm{IG}(a_2,b_2).
\end{eqnarray*}
In this context, the conditional posterior distribution of $\alpha^2$ given $\beta$ and $\mathbf{x}_n$ is
\begin{equation}\label{alpha2-post}
 \alpha^2|\beta,\mathbf{x}_n\sim\mathrm{IG}\left(\frac{n+1}{2}+a_2,\frac{1}{2}\sum_{i=1}^n\left(\frac{x_i}{\beta}+\frac{\beta}{x_i}-2\right)+b_2\right),
\end{equation}
and the marginal posterior distribution of $\beta$ given $\mathbf{x}_n$ satisfies
\begin{equation}\label{beta-post}
 \pi(\beta|\mathbf{x}_n)\propto\beta^{-(n+a_1+1)}\exp\left(\frac{b_1}{\beta}\right)\prod_{i=1}^n\left[\left(\frac{\beta}{x_i}\right)^{1/2}+\left(\frac{\beta}{x_i}\right)^{3/2}\right]\left[\frac{1}{2}\sum_{i=1}^n\left(\frac{x_i}{\beta}+\frac{\beta}{x_i}-2\right)+b_2\right]^{-\tfrac{n+1}{2}-a_2},
\end{equation}
which is not a known distribution \citep{Wang2016}. In this way, we use the random walk Metropolis-Hastings algorithm \citep{Metropolisetal53,Hastings70} to draw samples from the marginal posterior distribution of $\beta$ given $\mathbf{x}_n$. Using this sampling algorithm and the posterior distribution in~\eqref{alpha2-post} we may draw values from the joint posterior distribution of $\alpha^2$ and $\beta$. For a given $\mathbf{x}_n$, first we draw values of $\beta$ from (\ref{beta-post}) and given these values we draw values of $\alpha^2$ using (\ref{alpha2-post}). Note that the parameter of interest $\theta$ is the mean of the BS distribution and is a function of $\alpha^2$ and $\beta$. In order to obtain a random sample of the posterior distribution of $\theta$ given $\mathbf{x}_n$, we may draw values from the joint posterior of $\alpha^2$ and $\beta$, then apply~\eqref{eq:1} in each sampled pair of values.

\section{Optimal sample size}\label{sec:3}

We may approach the problem of determining the optimal sample size as a decision problem \cite[see][for example]{RaiffaSchlaifer1961,Parmigiani2009}. Given that $\theta$ is the parameter of interest, we specify a loss function $L(\theta, d_n)$ based on a sample $\mathbf{X}_n=(X_1,\ldots,X_n)$ and a decision function $d_n\equiv d_n(\mathbf{X}_n)$. For a given $n$ and depending on the adopted loss function, the action $d_n(\mathbf{x}_n)$ consists of the specification of one quantity (point inference case) representing an estimate for $\theta$, or two quantities (interval inference case) representing the lower and upper limits of a credible interval for $\theta$. Let $\pi$ be a prior distribution for the unknown parameter $\theta$ and $d_n$ a decision function; the Bayes risk is \citep{Parmigiani2009}
\begin{equation}\label{bayes-risk}
 r(\pi, d_n) \coloneqq \int_\Theta\int_{\mathcal{X}^n}L(\theta,d_n)g(\mathbf{x}_n|\theta)\pi(\theta)\dd\mathbf{x}_n\dd\theta,
\end{equation}
where $g(\cdot)$ is the sampling distribution for $\mathbf{X}_n$ given $\theta$, $\Theta$ is the parameter space, and $\mathcal{X}_n$ is the sample space. The decision $d_n^*$ that minimizes $r(\pi,d_n)$ among all the possible decisions $d_n$ is called the Bayes rule. In this context, we define the optimal sample size as the one that minimizes the total cost
\begin{equation*}
 TC(n) \coloneqq r(\pi, d_n^*) + C(n),
\end{equation*}
where $C(n)$ is the sampling cost function. Here, we take $C(n)=cn$, where $c$ is the per-unit cost for observing a unit in the population. Since it is not possible to compute $r(\pi,d_n^*)$ analytically, we use Monte Carlo simulations as an alternative to estimate $TC(n)$ for each $n$. Suppose that the order of the integration may be reverted in (\ref{bayes-risk}), then we have
\begin{equation*}
 r(\pi,d_n^*)=\int_{\mathcal{X}^n}\mathbb{E}[L(\theta,d_n^*)|\mathbf{x}_n]g(\mathbf{x}_n)\dd \mathbf{x}_n,
\end{equation*}
so that we may estimate the minimized Bayes risk through the posterior expected value of loss function applied to the Bayes rule $d_n^*$. This may be done as follows in the Algorithm~\ref{alg:1}.

\begin{algorithm}[b] % enter the algorithm environment
\SetAlgoLined
 \footnotesize
Set values for the hyperparameters;

Draw one value of $\alpha^2$ and one value of $\beta$ from the respective prior distributions, compute the square root of $\alpha^2$;
 
Given $\alpha$ and $\beta$, draw a value of $X_i$ from the $\mathrm{BS}(\alpha,\beta)$ using (\ref{stoch-eq}), for $i=1,\ldots,n$. This generates a sample $\mathbf{x}_n=(x_1,\ldots,x_n)$;
 
Given $\mathbf{x}_n$, draw a sample of size $N$ (as large as possible) from the joint posterior distribution of $\alpha^2$ and $\beta$ as explained in Section \ref{sec:2}. This generates values $(\alpha_j^2,\beta_j)$, $j=1,\ldots,N$; % PAREI AQUI
 
For $j=1,\ldots,N$, compute the posterior values $\theta_j$ using the generated values in Step 4 and (\ref{eq:1});
 
Obtain the respective Bayes rule $d_n^*$ using the sample of the posterior distribution of $\theta$ obtained in Step 5;
 
Use the values generated in Step 5 to compute an estimate of $\textrm{E}[L(\theta,d_n^*)|\mathbf{x}_n]$;
 
Repeat the Steps 1-7 $K$ times (as large as possible), this generates $K$ estimates of $\textrm{E}[L(\theta,d_n^*)|\mathbf{x}_n]$;
 
Take the average of the $K$ estimates obtained in Step 8, this is an estimate of $r(\pi, d_n^*)$.
 \caption{}
 \label{alg:1}
\end{algorithm}

After obtaining an estimate of $r(\pi, d_n^*)$ we add the respective cost sampling $cn$, which finally gives us an estimate of the total cost $TC(n)$ for a given $n$. We applied this procedure for a grid of plausible values of $n$. For example, if we set this grid of values as $n=2, 12,\ldots, 82, 92$, then we obtain an estimate for $TC(2), TC(12),\ldots, TC(82), TC(92)$. The choice of the grid of values is arbitrary and the smaller the span between its consecutive elements, the better to visualize the behavior of the total cost, but as we decrease this span the required computer processing power also increases, as well as the time to compute all these estimates. Thus, the choice of this grid must take into account all these settings.

In Step 4 of the Algorithm~\ref{alg:1}, when sampling from the marginal posterior distribution (\ref{beta-post}), we consider a burn-in of 500 iterations and a thinning of 20 with a final number of iterations of 500. We use these 500 iterations to compute an estimate of the Bayes risk. We inspect a trace and autocorrelation plot for a lower value of the grid used for $n$, we expect the same or better behavior as the $n$ increases in the grid. All the trace plots showed a random behavior around a value and in all the autocorrelation plots the autocorrelations for almost every lag were zero. In each value of $n$ in the grid, we estimate the Bayes risk ten times. 

\cite{Costa17} propose to fit the following curve to the grid of values of $n$ and the respective estimates of $TC(n)$, denoted by $tc(n)$
\begin{equation*}
 tc(n)=\frac{E}{(1+n)^G}+cn,
\end{equation*}
where $E$ and $G$ are parameters to be estimated. This curve may be linearized as a linear regression as follows
\begin{equation*}
 \log[tc(n)-cn]=\log E-G\log(1+n),
\end{equation*}
and the estimates of $E$ and $G$ may be computed by least squares. In this setting, the optimal sample size~($n_\mathrm{o}$) is the nearest integer closest to
\begin{equation*}
 \left(\frac{\widehat{E}\ \widehat{G}}{c}\right)^{1/(\widehat{G}+1)}-1,
\end{equation*}
\noindent
where $\widehat{E}$ and $\widehat{G}$ are, respectively, the least square estimates of $E$ and $G$.

\subsection{Loss functions}
We adopted four loss functions, the loss functions 1 and 2 may be used for point inference, {\it i.e.}, the decision $d_n$ provides a quantity representing an estimate for the parameter of interest $\theta$. The loss functions 3 and 4 may be used for interval inference, in this case, a decision provides two quantities, the lower (say, $a$) and the upper (say, $b$) limits of a credible interval for $\theta$. 

\subsubsection{Loss function 1 (L1)}

The first loss function is
\begin{equation*}
 L(\theta,d_n)=|\theta-d_n|,
\end{equation*}
which is known as the absolute loss function. For this loss function the Bayes rule $d_n^*$ is the median of the posterior distribution of $\theta$. Given a sample $\theta_j$, $j=1,\ldots,N$, of the posterior distribution of $\theta$, an estimate of $\mathbb{E}[L(\theta,d_n^*)|\mathbf{x}_n]$ may be obtained from $N^{-1}\sum_{j=1}^N|\theta_j-d_n^*|$.

\subsubsection{Loss function 2 (L2)}

Second, we consider the well-known quadratic loss function
\begin{equation*}
 L(\theta,d_n)=(\theta-d_n)^2,
\end{equation*}
for this loss function the Bayes rule $d_n^*$ corresponds to the posterior expected value of $\theta$ and in this case $\mathbb{E}[L(\theta,d_n^*)|\mathbf{x}_n]=\mathrm{Var}(\theta|\mathbf{x}_n)$. Given a sample $\theta_j$, $j=1,\ldots,N$, of the posterior distribution of $\theta$, an estimate of $\mathbb{E}[L(\theta,d_n^*)|\mathbf{x}_n]$ may be obtained from the respective sample variance.

\subsubsection{Loss function 3 (L3)}
The third loss function is
\begin{equation}\label{loss3}
L(\theta,d_n)=\rho\tau+(a-\theta)^++(\theta-b)^+,
\end{equation}
where $0<\rho<1$ is a weight, $\tau=(b-a)/2$ is the half-length of the desired interval, the function $x^+$ is equal to $x$ if $x> 0$ and equal to zero, otherwise. The smaller is $\tau$ the narrower the interval. The terms $(a-\theta)^+$ and $(\theta-b)^+$ are included to penalize intervals that do not contain the parameter of interest $\theta$. These terms are equal to zero if $\theta\in[a,b]$ and increase as $\theta$ moves away from the interval. Note that the loss function (\ref{loss3}) is a weighted sum of two terms, $\tau$ and $(a-\theta)^++(\theta-b)^+$, where the weights are $\rho$ and $1$, respectively. The Bayes rule $d_n^*$ corresponds to taking $a$ and $b$ as the quantiles of probabilities $\rho/2$ and $1-\rho/2$ of the posterior distribution of $\theta$. For more details see \cite{Riceetal08} or \cite{Costa17}. If we consider this loss function applied to the Bayes rule, we have
\begin{equation*}
 \mathbb{E}[L(\theta,d_n^*)|\mathbf{x}_n]=\mathbb{E}[\theta\delta_{\theta}(A_{b^*})|\mathbf{x}_n]-\mathbb{E}[\theta\delta_{\theta}(A_{a^*})|\mathbf{x}_n],
\end{equation*}
where $A_{b^*}=[b^*,\infty)$, $A_{a^*}=(0, a^*]$, $a^*$ and $b^*$ are the corresponding bounds of the Bayes rule $d_n^*$ and $\delta_\theta(\cdot)$ is the indicator function. Given a sample $\theta_j$, $j=1,\ldots,N$, of the posterior distribution of $\theta$, an estimate of $\mathbb{E}[L(\theta,d_n^*)|\mathbf{x}_n]$ may be obtained from $N^{-1}\sum_{j=1}^N[\theta_j\delta_{\theta_j}(A_{b^*})-\theta_j
\delta_{\theta_j}(A_{a^*})]$.

\subsubsection{Loss function 4 (L4)}

The last loss function is
\begin{equation*}
L(\theta,d_n)=\gamma\tau+(\theta-m)^2/\tau,
\end{equation*}
where $\gamma>0$ is a fixed constant and $m=(a+b)/2$ is the center of the credible interval. The first term involves the half-width of the interval and the second, the square of the distance between the parameter of interest $\theta$ and the center of the interval, which is divided by the half-width to maintain the same measurement unit of the first term. 

The weights attributed to each term are $\gamma$ and 1, respectively. If $\gamma<1$, we attribute the largest weight to the second term; if $\gamma>1$, the situation is reversed and if $\gamma=1$ the two terms have the same weight. For this loss function, the Bayes rule $d_n^*$ corresponds to the quantities which define the interval $[a^*,b^*]=[m^*-\mbox{SD}_\gamma, m^*+\mbox{SD}_\gamma]$, where $m^*=\mathbb{E}[\theta|\mathbf{x}_n]$ and $\mbox{SD}_\gamma=\gamma^{-1/2}[\mathrm{Var}(\theta|\mathbf{x}_n)]^{1/2}$. For more details see \cite{Riceetal08}, \cite{Parmigiani2009} or \cite{Costa17}. For this loss function, we have
\begin{equation*}
 \mathbb{E}[L(\theta,d_n^*)|\mathbf{x}_n]=2\gamma^{1/2}\sqrt{\mathrm{Var}(\theta|\mathbf{x}_n)}.
\end{equation*}
Given a sample $\theta_j$, $j=1,\ldots,N$, of the posterior distribution of $\theta$, an estimate of $\mathbb{E}[L(\theta,d_n^*)|\mathbf{x}_n]$ may be obtained from the respective sample variance and the previous equation.

\begin{figure}[!hb]
 \centering
 \includegraphics[width=13cm,height=8cm]{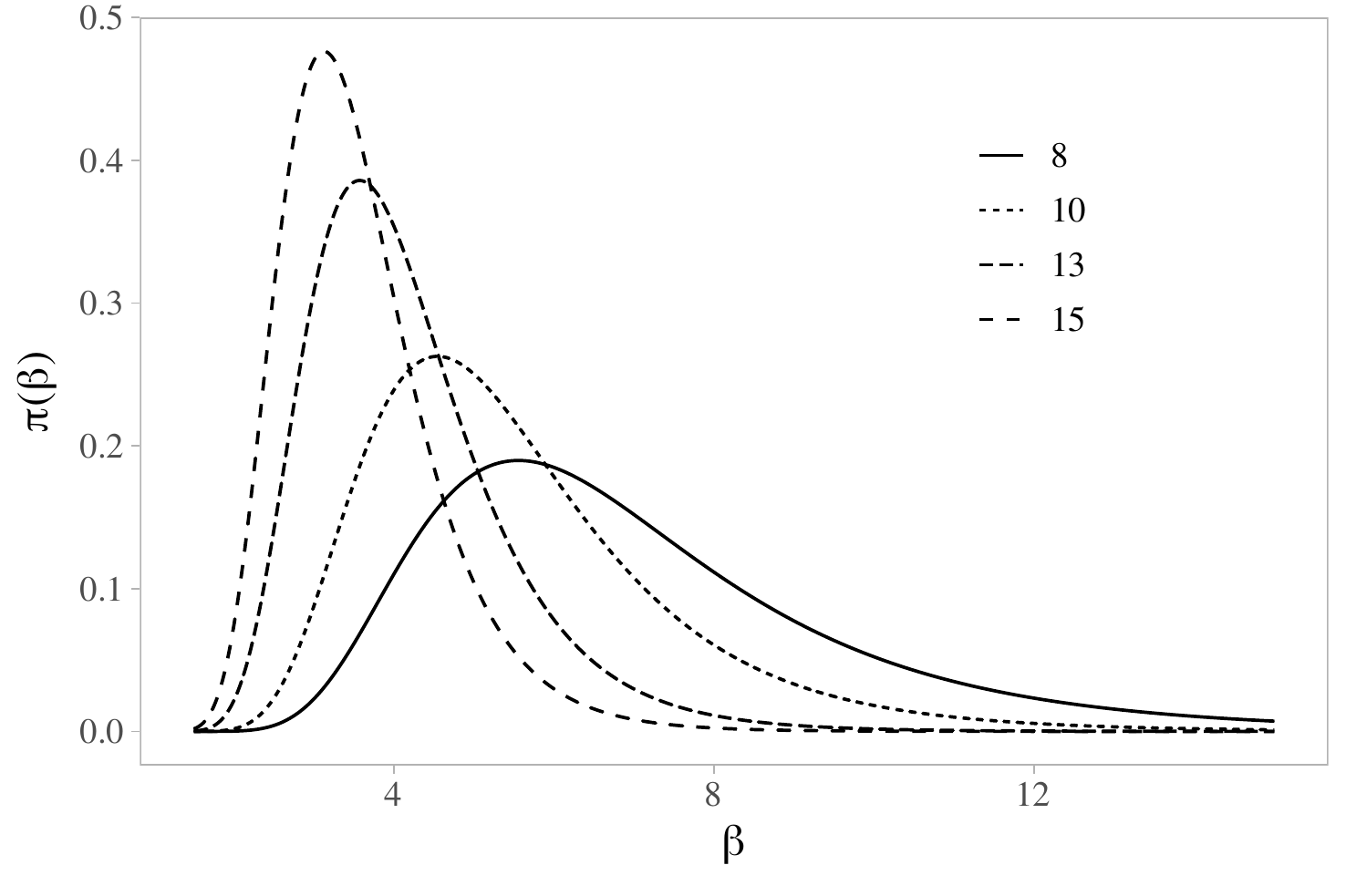}
 % figsim1.pdf: 0x0 px, 300dpi, 0.00x0.00 cm, bb=
 \caption{Density for different values of the hyperparameter $a_1$ ($b_1=50$) of the inverse gamma prior distribution for $\beta$.}
 \label{fig:sim}
\end{figure}

For the hyperparameters of the prior distribution of $\beta$, we consider $b_1=50$ and $a_1=8$, $10$, $13$ and $15$, with these values we have different degrees of prior information, see Figure~\ref{fig:sim}. For the prior distribution of $\alpha^2$, we set $a_2=a_1$ and $b_2=b_1$. We consider $c=0.001, 0.01$ and $0.1$ for the per-unit cost. For the loss function L3 we consider $\rho=0.01, 0.05$ and $0.10$, while for L4 we consider $\gamma=0.25, 0.50$ and $1.00$. For each combination of these values we compute the optimal sample size $n_\mathrm{o}$ for estimating $\theta$. The average acceptance rate for the Metropolis-Hastings algorithm in all these combinations was $\approx70\%$. Since the proposed methodology is based on simulation methods, we obtain $n_\mathrm{o}$ as triplicate and observe the difference between the three values. In Table~\ref{tab:sim1} we present the optimal sample sizes computed with these settings. 

An implementation of the proposed methodology is provided in the \proglang{R} language. The $n_\mathrm{o}$ may be obtained using the \proglang{R} package \proglang{samplesizeBS}~\citep[][]{Costa2020}. Also, the $n_\mathrm{o}$ may be obtained via the following link \url{https://santosneto.shinyapps.io/samplesizeBSapp/}, which also presents a graph with the fitted curve.

%\newcolumntype{s}{>{\columncolor[HTML]{AAACED}} p{3cm}}

%\arrayrulecolor[HTML]{DB5800}

\begin{sidewaystable}[h]
%\begin{table}[H]
   \renewcommand\arraystretch{1.1}
    \captionsetup{justification = centering}
	\caption{Optimal sample sizes $n_\mathrm{o}$ (in triplicate) for estimating the mean of the BS distribution via four different loss functions. }
	\resizebox{\textwidth}{!}{
\begin{tabular}{lSSS|SSS|SSS|SSS}
%\rowcolor{Tan!40}
\toprule
\multirow{2}*{{$\rho/\gamma$}}   & \multicolumn{3}{c|}{{$a_1=8$}}&\multicolumn{3}{c|}{{$a_1=10$}}&\multicolumn{3}{c|}{{$a_1=13$}}&\multicolumn{3}{c}{{$a_1=15$}}\\ \cmidrule(l){2-13} 
% \cmidrule(r){2-5}\cmidrule(r){6-9}\cmidrule(r){10-13}\cmidrule(r){14-17}
%\rowcolor{Tan!40} 
 & {$c=0.001$} & {$c=0.01$} & {$c=0.1$}
                & {$c=0.001$} & {$c=0.01$} & {$c=0.1$} 
                & {$c=0.001$} & {$c=0.01$} & {$c=0.1$}  
                & {$c=0.001$} & {$c=0.01$} & {$c=0.1$}  
\\ \midrule

%\cmidrule(r){2-5}\cmidrule(r){6-9}\cmidrule(r){10-13}\cmidrule(r){14-17}

%\rowcolor{Gainsboro!60} 
&\multicolumn{12}{c}{\bf L1 loss}\\ \cmidrule(l){2-13}
%\cmidrule(r){2-5}\cmidrule(r){6-9}\cmidrule(r){10-13}\cmidrule(r){14-17}
& 651&117&
23&436&80&
13&267&47&
9&210&36&6\\
%%%%%%%%%%%%%%%%%
& 641 &121 & 22 
& 429& 77&14 
& 267& 47&8 
& 209& 36&6  \\
%%%%%%%%%%%%%%%%%
&627 &140 &21  
&429 &77 &13 
&268 &47 &8 
&209 &37 &6  \\ \midrule
%\cmidrule(r){2-5}\cmidrule(r){6-9}\cmidrule(r){10-13}\cmidrule(r){14-17}
%\rowcolor{Gainsboro!60} 
&\multicolumn{12}{c}{\bf L2 loss}\\ \cmidrule(l){2-13}
%\cmidrule(r){2-5}\cmidrule(r){6-9}\cmidrule(r){10-13}\cmidrule(r){14-17}
&2096&641&176&1130&317&108 
&542&144 &33 
&381 &88 &21  \\
%%%%%%%%%%%%%%
&2129 &697 &200 
&1198 &326 &97 
&558 &138 &42 
&380 &89 &23  \\
%%%%%%%%%%%%%%%%%
&2075 & 622&218  
&1182 &292 &81 
&530 &139 &32 
&360 &89 &21  \\ \midrule
%\cmidrule(r){2-5}\cmidrule(r){6-9}\cmidrule(r){10-13}\cmidrule(r){14-17}
%\rowcolor{Gainsboro!60}
&\multicolumn{12}{c}{\bf L3 loss}\\
\cmidrule(l){2-13}
%\cmidrule(r){6-9}\cmidrule(r){10-13}\cmidrule(r){14-17}
                & 279 &53  &9  
                & 175 &31  &7 
                & 103 &18  &3  
                & 79 &14  &2 \\
                %%%%%%%%%%%%%      
{$\rho = 0.10 $}                &271  &54  & 10  
                &171  &31  &6 
                &106  &18  &3  
                & 79 &14  &2 \\
                %%%%%%%%%%%%%      
                & 284 &53  &10   
                & 168 &33  &5 
                & 103 &18  & 3 
                & 80 &14  &2 \\
 &187  &37  & 8
                &118  &22  & 4
                &70  &13  &2 
                &54  &9 &  \\
                %%%%%%%%%%%%          
{$\rho= 0.05 $}                &197  &37  &  8
                &121  &22  &4 
                &71  &12  &2  
                &55  &9  & \\
                %%%%%%%%%%%%%      
                &184  &40  &7   
                &121  &22  &4 
                &71  &13  &2  
                & 55 &9  & \\
&82&18  &  3
                &54  &9  &  
                &30  & 5 & 
                &22  &4  &  \\
                %%%%%%%%
{$\rho = 0.01 $}                &85  &19  &  3
                &52  &9  & 
                &29  &5  &  
                &22  &4  & \\
                %%%%%%%%%%%%%      
                &83  &18  &3   
                &49  &9  & 
                &30  &5  &  
                &22  &4  & \\ \midrule
%\cmidrule(r){2-5}\cmidrule(r){6-9}\cmidrule(r){10-13}\cmidrule(r){14-17}
%\rowcolor{Gainsboro!60} 
&\multicolumn{12}{c}{\bf L4 loss}\\
\cmidrule(l){2-13}
%\cmidrule(r){6-9}\cmidrule(r){10-13}\cmidrule(r){14-17}
                 &1461    &271   &  51  
                 &899    &171   &30     
                 &556    & 103  &18  
                 &441    &78   & 13   \\
                 %%%%%%%%%%%%%
 {$\gamma = 1.00$}&1472  &292  & 55  
                 &942   & 162 &31    
                 &561   &99  &18 
                 & 438  &78  &13   \\
                 %%%%%%%%%%%%%      
                &1460  &282  & 56  
                & 883 &162  & 30
                & 554 &101  & 18 
                &433  & 78 &13 \\
                 &1208  &203   &  39 
                 &684  &132   &23    
                 &427    &78   & 14  
                 &337   &59   &10   \\
                 %%%%%%%%%%%%
 {$\gamma = 0.50$}&1179   &201  &  38 
                 &690   & 130 &  23  
                 &434   &78  & 14
                 &335   &59  &10   \\
                 %%%%%%%%%%%%%      
                &1183  &213  & 42  
                &693  &134  & 24
                & 436 &80  & 14 
                &338  & 60 &10 \\
&796 &166   & 32 
                 &538  &106   &18  
                 &333    &59   & 10 
                 & 259   &46   &8  \\
                 %%%%%%%%%%%%%
  {$\gamma = 0.25$}&859   &171  & 30
                 &540   &99  & 19   
                 &331   &62  & 11
                 &260   &47  &  8 \\
                 %%%%%%%%%%%%%      
                &894  &167  & 32  
                &531  &101  & 18
                &333  &60  &10  
                &260  &46  &8 \\
\bottomrule
\end{tabular}
}
\label{tab:sim1}
%\end{table}
\end{sidewaystable}

\section{Discussion}\label{sec:4}

We propose a methodology to compute the optimal sample size for estimating the mean of the Birnbaum-Saunders distribution, a widely applied and studied distribution in several areas of science. We consider four different loss functions which allow to make both point and interval inference for the parameter of interest. 

An advantage of the proposed methodology is that the per-unit cost, represented by $c$, is explicitly taken into account. When the cost $c$ is fixed and $b_1=50$, the optimal sample size $n_\mathrm{o}$  decreases as the $a_1$ increases (or prior variance decreases) as expected, since in this case the prior knowledge increases as the $a_1$ increases. This occurs with all loss functions. For $b_1=50$ and $a_1$ fixed, the $n_\mathrm{o}$ also decreases as the $c$ increases; however, the total sampling cost decreases. For example, if we take the loss function L1, $a_1=8$ and $c=0.001$, the corresponding $n_\mathrm{o}$ is $651$ (Table \ref{tab:sim1}), which generates a total cost of $C(651)=0.001\times651=0.651$, whereas if we take $c=0.1$, the corresponding $n_\mathrm{o}$ is 23 (Table \ref{tab:sim1}), which generates a total cost of $C(23)=0.1\times23=2.3$. For the loss function L3, when $\rho$ increases the $n_\mathrm{o}$ also increases, if we consider $a_1$ e $c$ fixed. This makes sense because $\rho$ is the weight attributed to the term $\tau$ in L3, this term is related to the length of the credible interval and when we increases $\rho$ we expect longer credible intervals, consequently the probability of the respective interval decreases. The same is valid for $\gamma$ in the loss function L4, but in this case the decreasing of the respective credible interval is easily noted by the presence of the term $\gamma^{-\tfrac{1}{2}}$ in the expression of the respective Bayes rule. When $\gamma$ increases this term shrinks the length of the interval.

\begin{figure}[H]
 \centering
 \subfigure[][1st replica.]{\includegraphics[width=5.5cm,height=8.0cm]{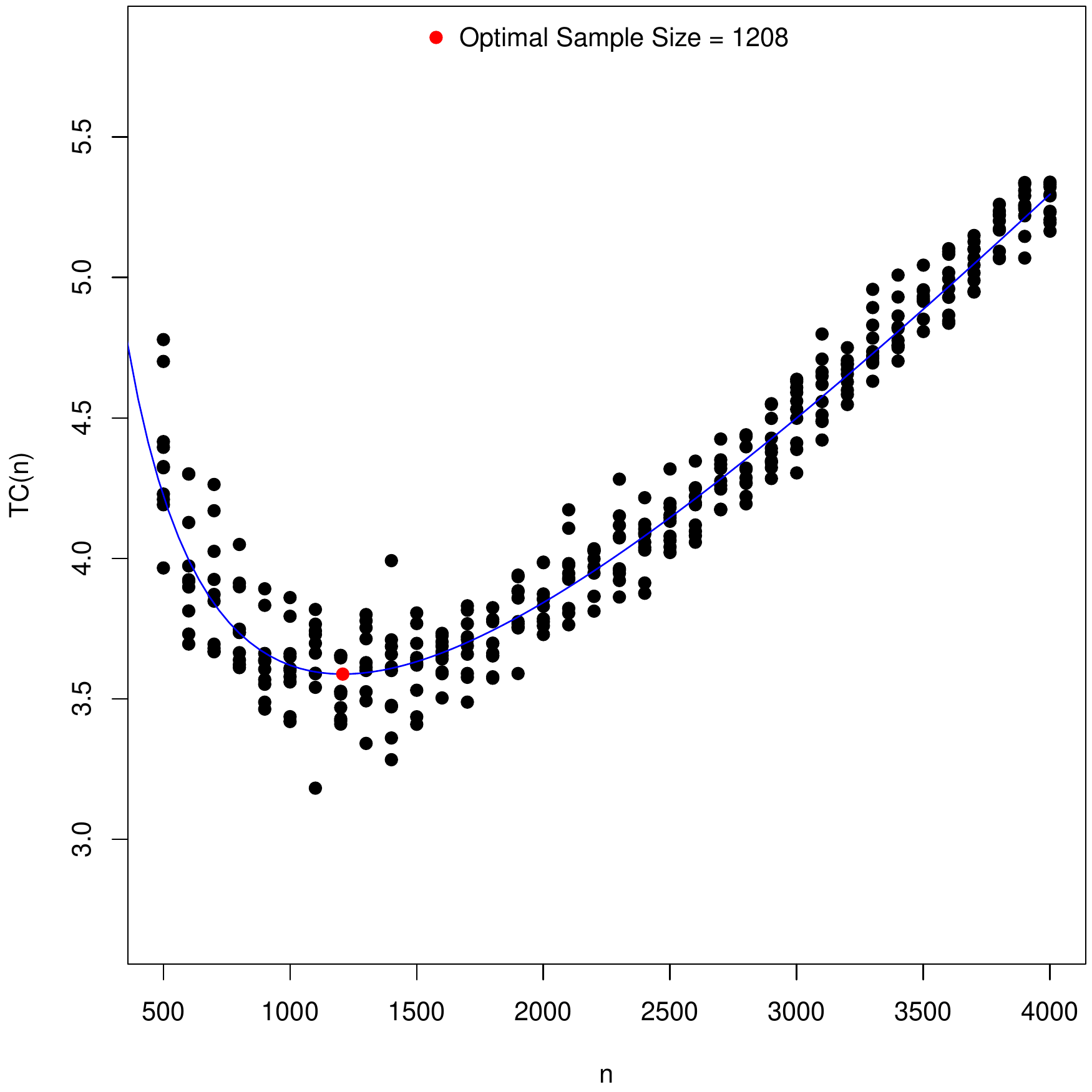}}
 \subfigure[][2nd replica.]{\includegraphics[width=5.5cm,height=8.0cm]{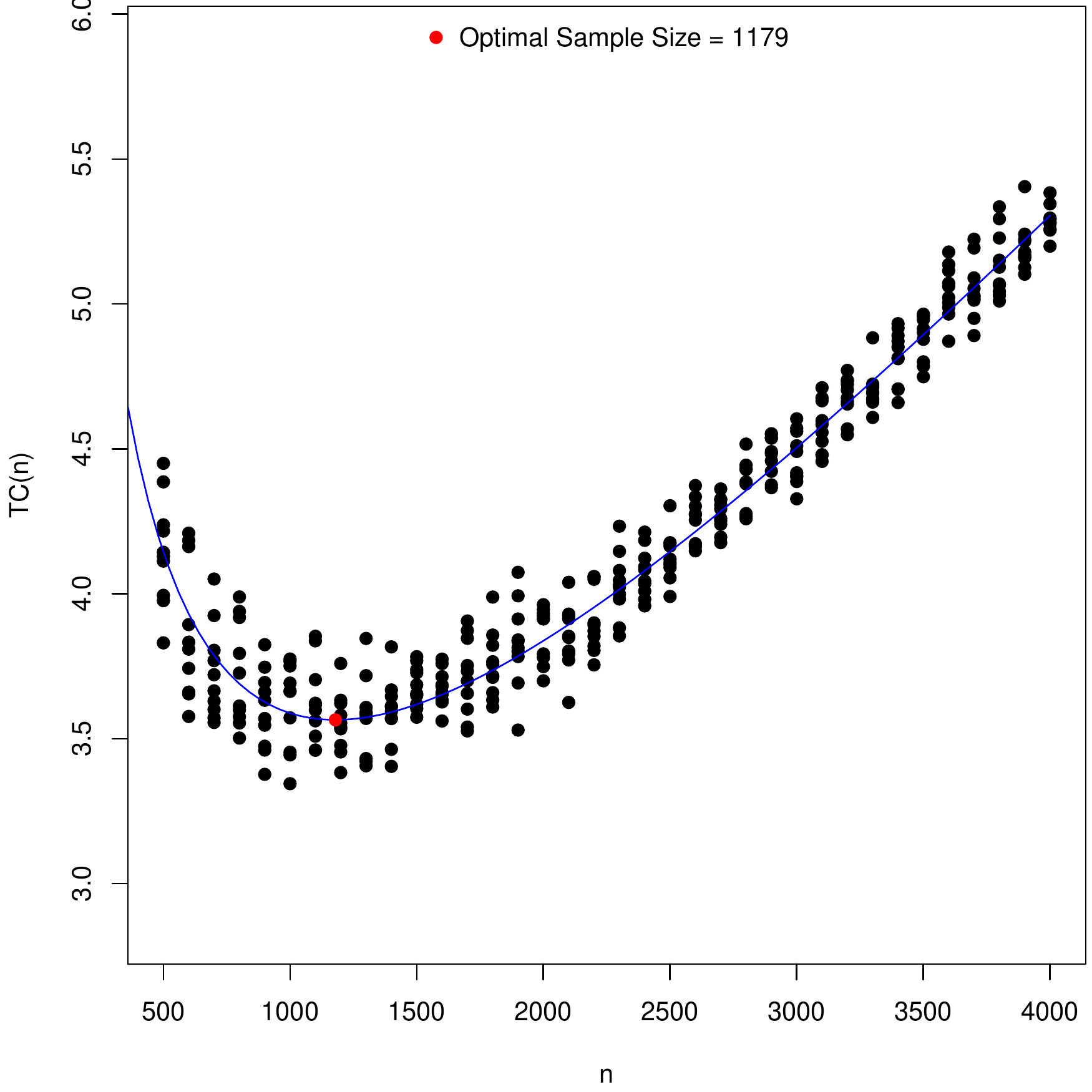}}
 \subfigure[][3rd replica.]{\includegraphics[width=5.5cm,height=8.0cm]{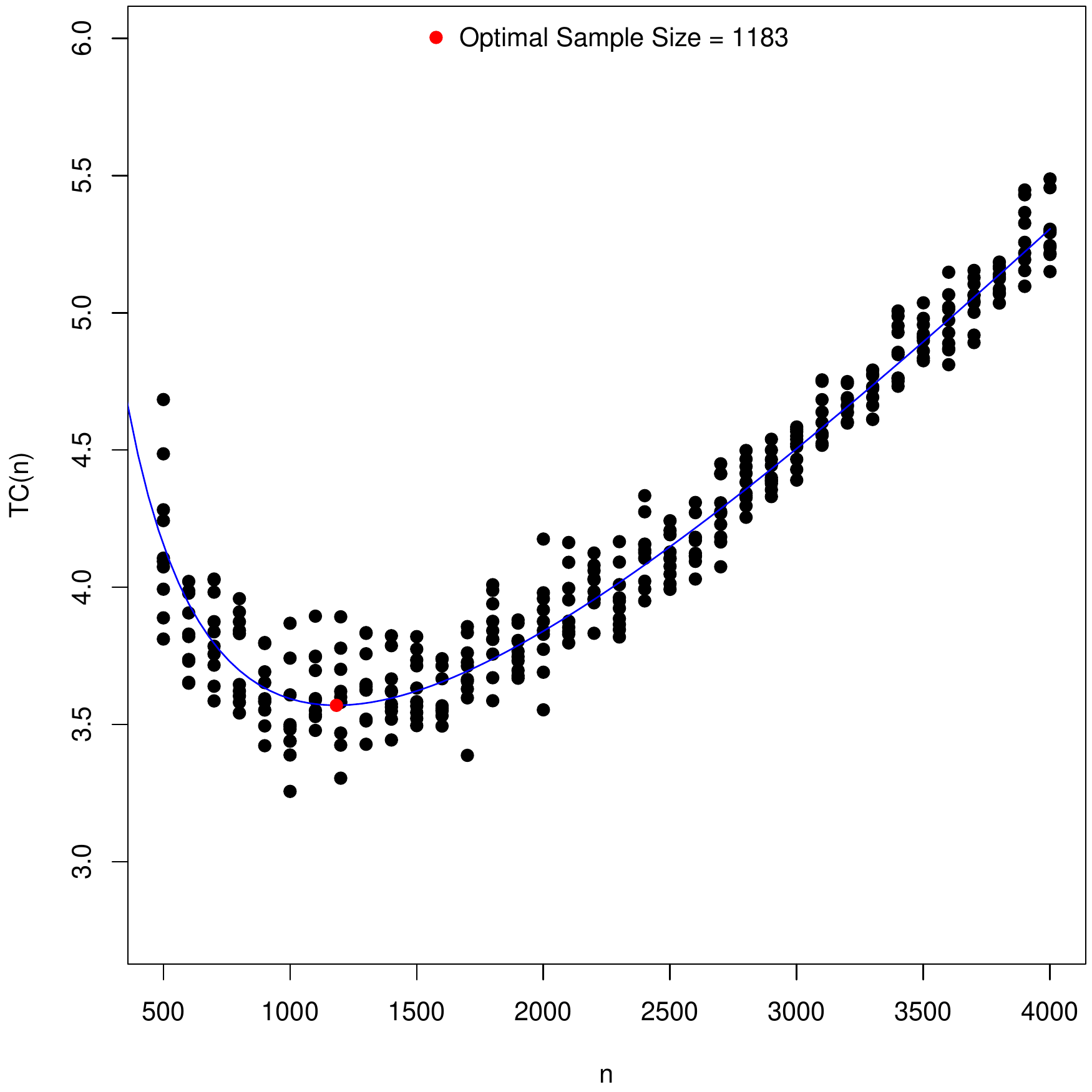}}
 \caption{Fitted curves with the respective optimal sample sizes obtained via the loss function L4 with $a_1=8$, $c=0.001$ and $\gamma= 0.50$.}
 \label{fig:curves}
 \end{figure}

Since the proposed methodology is based on simulations, we obtain the $n_\mathrm{o}$ in triplicate for each scenario of values of $a_1$, $c$, $\rho$ and $\gamma$. We observe that the largest discrepancies in the scenarios occur for $a_1=8$, these discrepancies decrease as the $a_1$ increases, or when the prior variance decreases. This also occurs when $c=0.001$ and/or when we consider the loss function L2. In general the discrepancy is close to zero, but if a large discrepancy occurs we suggest to inspect visually the graph of the fitted curves and take the value of $n_\mathrm{o}$ which corresponds to the best fit. However, if all the curves fit visually well, we suggest to use the median of the values obtained for $n_\mathrm{o}$. In our case we obtained the values of $n_\mathrm{o}$ in triplicate. For example, in Figure~\ref{fig:curves} under the loss function L4 with $a_1=8$, $c=0.001$ and $\gamma=0.50$ the values of $n_\mathrm{o}$ were 1208, 1179 and 1183. Since there is a discrepancy between these values and the fitting of the curves were visually well, in this case we suggest to use $n_\mathrm{o}=1183$.

Finally, note that we have no entry in Table \ref{tab:sim1} in some scenarios, which means that it is not worth sampling in these cases because the sampling cost outweighs the decreasing of the minimized Bayes risk. This was also observed by \cite{EtzioniKadane93} and \cite{IslamPettit14}. 

\section*{Acknowledgements}
 
Research carried out using the computational resources of the Center
for Mathematical Sciences Applied to Industry (CeMEAI) funded by Funda\c{c}\~ao de Amparo \`a Pesquisa do Estado de S\~ao Paulo (grant 2013/07375-0).

\clearpage

\bibliographystyle{authordate1} 
\bibliography{references}

\begin{thebibliography}{}

\bibitem[\protect\citename{Balakrishnan \& Kundu, }2019]{Balakrishnan2019}
Balakrishnan, N., \& Kundu, D. 2019.
\newblock Birnbaum-Saunders distribution: A review of models, analysis, and
  applications.
\newblock {\em Applied Stochastic Models in Business and Industry}, {\bf
  35}(1), 4--49.

\bibitem[\protect\citename{Birnbaum \& Saunders, }1969a]{bs69b}
Birnbaum, Z.~W., \& Saunders, S.~C. 1969a.
\newblock Estimation for a family of life distributions with applications to
  fatigue.
\newblock {\em Journal of Applied Probability}, {\bf 6}(2), 328--347.

\bibitem[\protect\citename{Birnbaum \& Saunders, }1969b]{bs69}
Birnbaum, Z.~W., \& Saunders, S.~C. 1969b.
\newblock A new family of life distributions.
\newblock {\em Journal of Applied Probability}, {\bf 6}(2), 319--327.

\bibitem[\protect\citename{Bourguignon {\em et~al.\ }\relax,
  }2020]{Bourguignonetal20}
Bourguignon, M., Lee~Ho, L., \& Fernandes, F.~H. 2020.
\newblock Control charts for monitoring the median parameter of
  Birnbaum-Saunders distribution.
\newblock {\em Quality and Reliability Engineering International}, {\bf 36}(4),
  1333--1363.

\bibitem[\protect\citename{Costa, }2017]{Costa17}
Costa, E.~G. 2017.
\newblock {\em Tamanho amostral para estimar a concentra\c c\~ao de organismos
  em \'agua de lastro: uma abordagem bayesiana}.
\newblock Ph.D. thesis, {Departamento de Estat\'istica, Universidade de S\~ao
  Paulo}, S\~ao Paulo.
\newblock In Portuguese. DOI: 10.11606/T.45.2018.tde-05072018-164225.

\bibitem[\protect\citename{Costa \& Santos-Neto, }2020]{Costa2020}
Costa, Eliado~G., \& Santos-Neto, Manoel. 2020.
\newblock {\em samplesizeBS: Bayesian sample size in a decision-theoretic
  approach for the Birnbaum-Saunders}.
\newblock \url{www.github.com/santosneto/samplesizeBS}.
\newblock R package version 0.1.5.

\bibitem[\protect\citename{De~Santis \& Gubbiotti, }2016]{DeSantisGubbiotti16}
De~Santis, F., \& Gubbiotti, S. 2016.
\newblock A decision-theoretic approach to sample size determination under
  several priors.
\newblock {\em Applied Stochastic Models in Business and Industry,}.
\newblock {doi: 10.1002/asmb.2211}.

\bibitem[\protect\citename{Etzioni \& Kadane, }1993]{EtzioniKadane93}
Etzioni, R., \& Kadane, J.~B. 1993.
\newblock Optimal experimental design for another's analysis.
\newblock {\em Journal of the American Statistical Association}, {\bf 88}(424),
  1404--1411.

\bibitem[\protect\citename{Hastings, }1970]{Hastings70}
Hastings, W.~K. 1970.
\newblock Monte {C}arlo sampling methods using {M}arkov chains and their
  applications.
\newblock {\em Biometrika}, {\bf 57}(1), 97--109.

\bibitem[\protect\citename{Islam \& Pettit, }2012]{Islam2012}
Islam, A. F. M.~S., \& Pettit, L.~I. 2012.
\newblock Bayesian Sample Size Determination Using Linex Loss and Linear Cost.
\newblock {\em Communications in Statistics - Theory and Methods}, {\bf 41}(2),
  223--240.

\bibitem[\protect\citename{Islam \& Pettit, }2014]{IslamPettit14}
Islam, A. F. M.~S., \& Pettit, L.~I. 2014.
\newblock Bayesian sample size determination for the bounded linex loss
  function.
\newblock {\em Journal of Statistical Computation and Simulation}, {\bf 84}(8),
  1644--1653.

\bibitem[\protect\citename{Leiva, }2016]{leivabook}
Leiva, V. 2016.
\newblock {\em The Birnbaum-Saunders distribution}.
\newblock New York: Academic Press.

\bibitem[\protect\citename{Metropolis {\em et~al.\ }\relax,
  }1953]{Metropolisetal53}
Metropolis, N., Rosenbluth, A.~W., Rosenbluth, M.~N., Teller, A.~H., \& Teller,
  E. 1953.
\newblock Equation of state calculations by fast computing machines.
\newblock {\em Journal of Chemical Physics}, {\bf 21}(6), 1087--1092.

\bibitem[\protect\citename{Parmigiani \& Inoue, }2009]{Parmigiani2009}
Parmigiani, G., \& Inoue, L. 2009.
\newblock {\em Decision theory: principles and approaches}.
\newblock New York: John Wiley \& Sons.

\bibitem[\protect\citename{Raiffa \& Schlaifer, }1961]{RaiffaSchlaifer1961}
Raiffa, H., \& Schlaifer, R. 1961.
\newblock {\em Applied statistical decision theory}.
\newblock Boston: Harvard University Press.

\bibitem[\protect\citename{Rice {\em et~al.\ }\relax, }2008]{Riceetal08}
Rice, K.~M., Lumley, T., \& Szpiro, A.~A. 2008.
\newblock {\em Trading bias for precision: decision theory for intervals and
  sets}.
\newblock \url{http://www.bepress.com/uwbiostat/paper336}.
\newblock Working Paper 336, UW Biostatistics.

\bibitem[\protect\citename{Sahu \& Smith, }2006]{Sahu2006a}
Sahu, S.~K., \& Smith, T. M.~F. 2006.
\newblock A Bayesian method of sample size determination with practical
  applications.
\newblock {\em Journal of the Royal Statistical Society: Series A (Statistics
  in Society)}, {\bf 169}(2), 235--253.

\bibitem[\protect\citename{Wang {\em et~al.\ }\relax, }2016]{Wang2016}
Wang, M., Sun, X., \& Park, C. 2016.
\newblock Bayesian analysis of Birnbaum-Saunders distribution via the
  generalized ratio-of-uniforms method.
\newblock {\em Computational Statistics}, {\bf 31}(1), 207--225.

\end{thebibliography}

\end{document}